\documentclass[twocolumn,superscriptaddress,prx]{revtex4-1}%
\usepackage{mathrsfs}
\usepackage{amsmath}
\usepackage{amsfonts}
\usepackage{amssymb}
\usepackage{amsthm}
\usepackage{bm}
\usepackage{bbm}
\usepackage{float}
\usepackage{graphicx}
\usepackage{subfigure}
\usepackage[outercaption]{sidecap}
\usepackage{color}
\usepackage{hyperref}
\usepackage{cleveref}
\providecommand{\U}[1]{\protect\rule{.1in}{.1in}}
\DeclareMathAlphabet\mathbfcal{OMS}{cmsy}{b}{n}
\usepackage{extarrows}

\begin{document}

\title[Anomalous quantum interference effects in graphene SNS junctions]{Anomalous quantum interference effects in graphene SNS junctions due to strain-induced gauge fields}

\author{Hadi Khanjani}
\address{Department of Physics, Institute for Advanced Studies in Basic Sciences (IASBS), Zanjan 45137-66731, Iran}

\author{Ali G. Moghaddam}
\address{Department of Physics, Institute for Advanced Studies in Basic Sciences (IASBS), Zanjan 45137-66731, Iran}
\address{Research Center for Basic Sciences \& Modern Technologies (RBST), Institute for Advanced Studies in Basic Science (IASBS), Zanjan 45137-66731, Iran}

\begin{abstract}
We investigate the influence of gauge fields induced by strain on the supercurrent passing through the graphene-based Josephson junctions. We show in the presence of a constant pseudomagnetic field ${\bf B}_S$ originated from an arc-shape elastic deformation, the Josephson current is monotonically enhanced. This is in contrast with the oscillatory behavior of supercurrent (known as Fraunhofer pattern) caused by real magnetic fields passing through the junction. The absence of oscillatory supercurrent originates from the fact that strain-induced gauge fields have opposite directions at the two valleys due to the time-reversal symmetry. Subsequently there is no net Aharonov-Bohm effect due to ${\bf B}_S$ in the current carried by the bound states composed of electrons and holes from different valleys. On the other hand, when both magnetic and pseudomagnetic fields are present, Fraunhofer-like oscillations as function of the real magnetic field flux are found. We find that the Fraunhofer pattern and in particular its period slightly change by varying the strain-induced gauge field as well as the geometric aspect ratio of the junction. Intriguingly, the combination of two kinds of gauge fields results in two  special fingerprint in the local current density profile: (i) strong localization of the Josephson current density with more intense maximum amplitudes; (ii) appearance of the inflated vortex cores - 
finite regions with almost diminishing Josephson currents - which their sizes increases by increasing ${\bf B}_S$. These findings reveal unexpected interference signatures of strain-induced gauge fields in graphene SNS junctions and provide unique tools for sensitive probing of the pseudomagnetic fields.    
\end{abstract}



\maketitle
\section{Introduction}
Gauge fields are among the most fundamental cornerstones of modern theoretical physics. They lie in the heart of standard model which successfully describes many aspects of the elementary particles \cite{weinberg-book}. On the other hand, their success is widespread in many branches of physics, particularly in the context of condensed matter physics \cite{kleinert-1989,fradkin-book,wen-book} and very recently in ultracold atoms \cite{jaksch-2003,lin-2009, dalibard-2011,goldman-2014}. Historically, the concept of gauge fields has been widely used in various areas of condensed matter physics, including the theory of phase transitions, glasses, topological defects \cite{kleinert-1989}, and importantly the effects related to the Berry phase \cite{niu-2009}. However, a breakthrough came out after the synthesis of graphene, a two-dimensional (2D) atomic monolayer, in which gauge fields emerge from elastic deformations\cite{guinea-1992,ando-prb-2002,manes-prb-2007,morozov-prl-2006, morpurgo-prl-2006,castroneto-rmp-2009,guinea-rep}. 
The emergence of gauge fields in graphene originates from its relativistic spectrum and the fact that the main effect of smooth strain is to relocate and deform the Dirac cones in the corners of Brillouin zone \cite{castroneto-rmp-2009,guinea-rep}. Subsequently, the dynamics of electrons in strained graphene is reminiscent of motion at the presence of a magnetic vector potential ${\bf A}$ but with a special property of having opposite signs at the two Dirac points, to preserve time-reversal symmetry (TRS) \cite{guinea-prb-2009,katsnelson-natphys}. The existence of such pseudomagnetic fields has been experimentally validated by some experiments, especially using spectroscopy of Landau level at the vicinity of a highly strained nanobubble which has shown very large fields exceeding a few hundred Tesla \cite{crommie-science-2010}. 
\par
One of the very intriguing aspects of a gauge potential ${\mathbfcal A}$,
is attributed to Aharonov and Bohm who showed a charged particle propagating over a path ${\cal C}$ will acquire an extra phase $\varphi_{\rm AB}\propto\int_{\cal C} {\mathbfcal A} \cdot d{\bf r}$ in its wave-function \cite{aharonov-bohm}.
Such a purely quantum mechanical effect leads to the interference phenomena which have been experimentally manifested in various systems in the past decades \cite{popescu85}. In solid state systems, the superconductor (S) with robust phase coherence provides a very promising building block for Aharonov-Bohm (AB) interference experiments \cite{yamada86}. Flux quantization in a superconducting loop and Fraunhofer diffraction pattern in Josephson current are famous examples of macroscopic quantum interferences as intimate manifestations of the AB effect \cite{Tinkham,doll61,deaver61,rowell-1963,anderson-rowell,jaklevic}. These phenomena are extensively used in very sensitive measurements of magnetic fields, particularly in the so-called superconducting quantum interference devices \cite{clarke2006squid}. Of particular interest, when a magnetic flux $\Phi_B$ is imposed 
to the non-superconducting or normal (N) region sandwiched between two superconductors, the critical supercurrent $I_c$ shows a diffraction pattern by varying $\Phi_B$. In addition, the local current density inside the N region reveals the so-called Josephson vortices which are governed by quantum mechanical interferences in contrast to the 
Abrikosov vortex lattice where electrostatics plays a major physical role \cite{roditchev2015,blatter-99,ostroukh-2016}. 
\par
The aim of current work is to investigate the influence of strain-induced gauge fields on the Josephson current and vortices in superconductor-graphene-superconductor (SGS) junctions. Over the last decade, many theoretical works have been focused on SGS and other graphene based superconducting heterostructures and found various peculiar and unexpected behaviors \cite{beenakker-2006,titov-2006,sengupta-2006,moghaddam-2007,
linder-2008,moghaddam-2008,beenakker-2008,
cserti-2010}. Most of the these features originate from Dirac dispersion of quasiparticles as well as 2D nature of graphene as pointed out first by Beenakker \cite{beenakker-2006,beenakker-2008}.
Interestingly, various experiments have even outpaced theory owe to the impressive techniques in fabrication of high quality graphene based devices and very good contacts with various superconducting materials \cite{heersche-2007,andrei-2008, girit-2009,bouchiat-2009,borzenets-2011,lee-prl-2011,
coskun-2012,mizuno-2013,choi-2013,
vandersypen-2015,benshalom-2015,morpurgo-2014,yacoby-natphys,kim-natphys}. Some recent works have studied exhaustively the Fraunhofer pattern in SGS junctions \cite{cserti-2016,vandersypen-2015,benshalom-2015}. Nevertheless, the effect of gauge fields and particularly the corresponding quantum interferences in the SGS systems has remained almost unexplored so far. 
\par
Here, based on a semiclassical framework, we show under a general gauge potential $\mathbfcal{A}$, two different phases $\chi$ and $\chi'$ are gained by the quasiparticles wave-functions. The first phase $\chi$ depends on the flux of gauge field resembling the well-known AB effect, while $\chi'$ acting as a relative phase between the two components of the Dirac spinor depends on the circumference of the quasiparticles trajectories. As a key finding, it is demonstrated that Josephson current is enhanced at the presence of a uniform strain-induced pseudomagnetic field ${\bf B}_S$. In fact oscillatory Fraunhofer-like pattern of supercurrent is absent since the strain-induced AB phases gained by the electron and hole making a bound states, cancel each other. Nevertheless, when a real magnetic field ${\bf B}$ is imposed to the junction beside the pseudomagnetic field, Fraunhofer pattern establishes by its variation. Further investigation of the combined effect of real and strain-induced fields, reveals that Josephson vortices are strongly influenced by the presence of ${\bf B}_S$. In one hand, the gauge-fields presence further localizes the vortex pattern. On the other hand, new vortex cores appear and inflate significantly at the presence of gauge fields which means finite regions of almost vanishing supercurrent emerge having very sharp boundaries with nonvanishing supercurrent regions.  
\par
The paper is organized as follows. After the introduction, in Sec. \ref{model} the basic model, semiclassical framework for the calculations, and the main relations for the supercurrent are presented. Then, in Sec. \ref{results} the Josephson current and its local density dependence on the gauge fields are shown followed by a discussion over the importance of the results and their experimental relevance. Finally, Sec. \ref{conc} is devoted to the concluding remarks. 
\section{Model and basic formalism}\label{model}
\begin{figure}[tp!]
\centering
\includegraphics[width=0.95\linewidth]{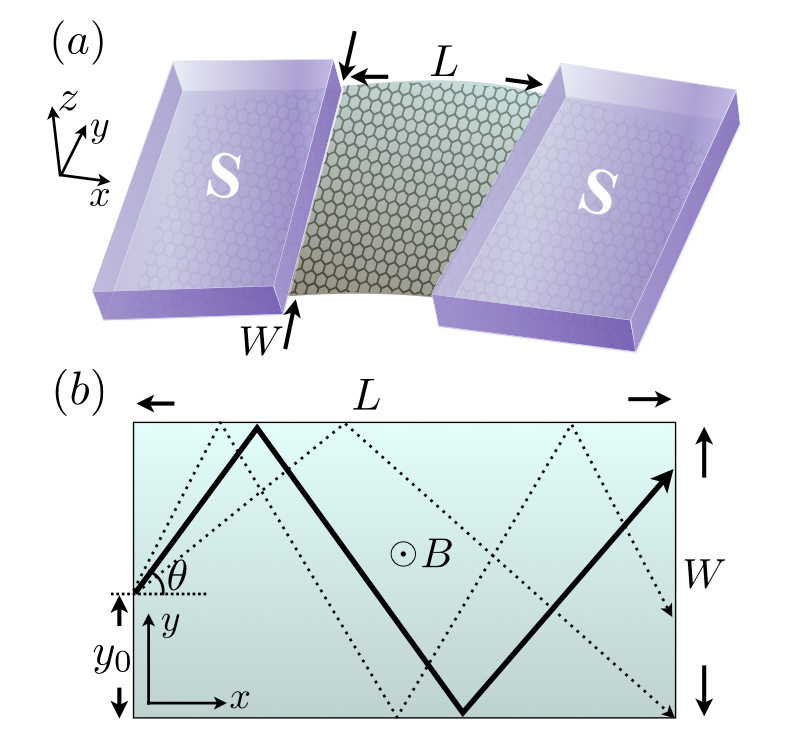}
\caption{(Color online) (a) Schematic of the SGS Josephson junction with strained graphene at the middle. (b) The classical trajectories of Dirac excitations inside the normal graphene. 
The gauge fields are weak enough so that they cannot deflect the trajectories and only phase effects corresponding to them can come into play. The geometric parametrization of the trajectories and the normal region are shown. }
\label{fig1}
\end{figure}

Our model consists of an SGS junction where two superconducting electrodes are deposited over graphene with distance $L$ from each other as depicted in Fig. \ref{fig1}(a). The normal graphene region is imposed to an arc-shaped strain or alternatively deformation with triangular symmetry which can lead to a uniform pseudomagnetic field ${\bf B}_S=B_S {\bf z}$ acting on the Dirac quasiparticles \cite{guinea-prb-2009,katsnelson-natphys}. The strain-induced field and corresponding gauge potential ${\bf A}_S$ (${\bm \nabla}\times {\bf A}_S \equiv {\bf B}_S$) have opposite signs at the vicinity of the two Dirac points $K$ and $K'$ at the corners of hexagonal first Brillouin zone \cite{guinea-rep}. The underlying deep reason for the sign change originates from the facts that strain and any geometric deformation do not break TRS, and the two K-points or valleys are connected by time reversal operator ($K \xlongleftrightarrow{\Theta} K'$). Putting together the low energy Hamiltonian of normal graphene around Dirac points and at the presence of both strain-induced gauge potentials and real magnetic fields (${\bf B}={\bm \nabla}\times {\bf A}$) can be written as,
\begin{equation}
{\cal \check {H}}({\bf A,A_{S}}) = v_{F}\begin{pmatrix} \hat{\bm\sigma}\cdot{\bf \Pi}_K &0 \\
0 & \hat{\bm\sigma}^\ast\cdot{\bf \Pi}_{K'}
\end{pmatrix},
\end{equation}
with ${\bf \Pi}_{K,K'}={\bf p}+e({\bf A}\pm{\bf A}_S)$ indicating the kinematical momenta at the vicinity of two valleys and $\hat{\bm \sigma}=(\sigma_x,\sigma_y)$ the Pauli matrices in the sublattice or pseudospin space. Here the Fermi velocity is denoted by $v_F$ and ${\bf p}=(p_x,p_y)$ is the 2D momentum. Moreover, for the sake of clarity, any quantity in 
2D space of sublattices and four-dimensional space composed of valley and sublattice subspaces are labeled with \emph{hat} and \emph{check} signs, respectively.
\par
Inside the parts of graphene covered by superconducting electrodes, an effective pair potential of the form $\check{\Delta}=\Delta\,\check{\mathbb I}$ is induced via proximity effect. We assume a phase difference $\phi$ between two superconductors with pairing functions $\Delta_{L,R}=\Delta_0\exp(\pm i\phi/2)$, to drive the Josephson current. The coupling of electron and hole excitations ($\check{\Psi}_{e,h}$) inside these superconducting graphene regions are governed by the following Bogoliubov-de Gennes (BdG) equation,
\begin{equation}
\begin{pmatrix}
\check{\cal H}-\mu&\check{\Delta}\\
\check{\Delta}&\mu-\check{\Theta} {\check{\cal H}} \check{\Theta}^{-1}
\end{pmatrix}\begin{pmatrix}
\check{\Psi}_{e}\\
\check{\Psi}_{h}
\end{pmatrix}=\varepsilon\begin{pmatrix}
\check{\Psi}_{e}\\
\check{\Psi}_{h}
\end{pmatrix},
\end{equation}
in which $\mu$ and $\varepsilon$ indicate the chemical potential and the energy of the Bogoliubov quasiparticles, respectively. The time-reversed Hamiltonian corresponding to the hole-type excitations follows $\check{\Theta} {\check{\cal H}}({\bf A},{\bf A}_S) \check{\Theta}^{-1}={\check{\cal H}}(-{\bf A},{\bf A}_S)$, Since under the act of time-reversal operator $\check{\Theta}$ the real magnetic fields change sign while the strain-induced gauge fields remain the same.
\subsection{Constructing a semiclassical framework}
In what follows we will construct a semiclassical  
picture for the propagation of massless Dirac particles at the presence of a general Abelian gauge potential ${\mathbfcal A}$ with contributions from both real magnetic field and geometric deformations. 
The starting point is to write the massless Dirac equation in the squared form, 
\begin{equation}
v_F^2[\hat{\bm \sigma}\cdot({\bf p}+e {\mathbfcal A})]^2\hat{\Psi}=E^2 \hat{\Psi}
\end{equation}
Assuming the gauge potentials are weak enough, we only keep the linear terms in 
${\mathbfcal A}$ and corresponding gauge field ${\cal B}={\bf z}\cdot ({\bm \nabla}\times {\mathbfcal A})$ which results in, 
\begin{equation}
v_F^2({\bf p}^2+2e {\mathbfcal A}\cdot {\bf p} + \hbar e{\cal B}\hat{\sigma}_z )\hat{\Psi}=E^2 \hat{\Psi}
\end{equation}
The main step towards the semiclassical framework can be passed by considering the wave-function as $\hat{\Psi}({\bf r})=e^{i{\bf k}\cdot{\bf r}}\hat{\psi}({\bf r})$ composed of a plane-wave corresponding to the energy $E=\hbar v_F |{\bf k}|$ and a slowly varying envelope function $\hat{\psi}({\bf r})$.
Then we can ignore second order spatial derivative of  $\hat{\psi}$ as well as the term proportional to ${\mathbfcal A}\cdot{\bm \nabla}\hat{\psi}$ provided by two legitimate assumptions $|\nabla^2 \hat{\psi}| \ll |{\bf k}\cdot{\bm\nabla} \hat{\psi}|$ and $|e {\mathbfcal A}| \ll \hbar k$ when energy of the excitation with respect to Dirac point is high enough. 
As a result we arrive in the following first-order equation of motion for the envelope function $\hat{\psi}({\bf r})$,
\begin{equation}
\left(-2i \hbar^2 {\bf k}\cdot {\bm \nabla} 
+2\hbar e  {\bf k} \cdot \mathbfcal{ A} +\hbar e{\cal B}\hat{\sigma}_z\right)\hat{\psi}=0,
\end{equation}
corresponding to the semiclassical trajectories determined by the direction of the wave-vector ${\bf n}={\bf k}/k_F$ (Throughout the paper only excitations at the vicinity of the Fermi level are considered). It must be mentioned that under our assumption the trajectories are straight lines and the gauge fields are too weak to bend them. More precisely, the cyclotron radius $R_{\rm cyc}\sim \hbar k_F /e {\cal B}$ over which the bending of a trajectory take place is much larger than length $L$ and width $W$ of the normal graphene region. Figure \ref{fig1}(b) shows a few examples of classical trajectories started from one SG interface and ended at another. It is clear that the trajectories are mirror reflected at the edges of the graphene sheet along $x$ direction where hard-wall boundary conditions must be satisfied. The trajectories can be fully determined by their  initial position at the interface denoted by $y_0$ and the angle $\theta$. Then each trajectory can be parametrized with a single parameter $\ell$ indicating the distance from initial point along the trajectory. This leads to the following equation, 
\begin{equation}
\partial_{\ell}\hat{\psi}=-\frac{ie}{\hbar}
\left( {\bf n}\cdot \mathbfcal{ A} +\frac{\cal B}{2k_F} \hat{\sigma}_z
\right)\hat{\psi},\label{evolution}
\end{equation}
which has the form of evolution equation and provides a cornerstone for our semiclassical description of Dirac equation at the presence of gauge fields.
\par
Considering any trajectory ${\cal C}$, the above semiclassical equation can be (formally) integrated as,
$\hat{\psi}_{\cal C}=\exp[-i(\chi+\chi'\hat{\sigma}_z)] \hat{\psi}_0$, with two gauge-induced phase factors,
\begin{equation}\label{chi-chip}
\chi= \frac{\pi}{\Phi_0}\int_{\cal C} d\ell~
{\bf n}\cdot \mathbfcal{ A}~~,~~~~ 
\chi'= \frac{\pi}{2k_F\Phi_0}
\int_{\cal C} d\ell~ {\cal B}~,
\end{equation}
where $\Phi_0=\pi \hbar/e$ is the superconducting magnetic flux quantum. While the first phase $\chi$ is responsible for the conventional AB effect, the second one $\chi'$ is quite different and acts as a respective phase between the components of the spinor $\hat{\psi}$ rather than being an overall phase. In particular, while Stokes' theorem guarantees that for a closed path, $\chi$ depends on the area enclosed by the closed trajectory $\cal{C}$ and more precisely the flux, however, $\chi'$ is a function of circumference. Another point must be mentioned is that the AB phase $\chi$ for the closed paths and $\chi'$ for any trajectory (open or closed) are clearly gauge invariant, as expected. Considering a constant gauge field ${\cal B}_z$ and choosing a Landau gauge $\mathbfcal{A}=-{\cal B}_z y \,{\bf x}$, for a given trajectory determined by $y_0$ and $\theta$, the two phase factors obey the following relations, 
\begin{equation} \label{phases}
\chi=\frac{\pi S_{y_0,\theta}}{L W}\frac{ \Phi_{\cal B}}{\Phi_0}
~~,~~~
\chi'=\frac{\pi}{2 k_F W\cos\theta}\frac{ \Phi_{\cal B}}{\Phi_0}~.
\end{equation}
Here, $\Phi_{\cal B}=LW{\cal B}$ is the gauge flux passing through the nonsuperconducting graphene region, and $S_{y_0,\theta}$ indicates the area enclosed between the trajectory and the lower parts of the normal region boundaries. The exact form of the area will be shown in \ref{appendix} with details of the calculations. It will be very useful for the discussion in next section, if we rewrite the condition for having straight trajectories versus the scaled flux $\Phi_{\cal B}/\Phi_0$. One can simply check that the large cyclotron radius is equivalent to $\pi \Phi_{\cal B}/\Phi_0 \lesssim {\rm min}(k_FL,k_FW)$ where both $k_FL$ and $k_FW$ are very large to guarantee the validity of semiclassical picture. Therefore, while for the angles $\theta\to\pm\pi/2$ the phase $\chi'$ could be very large, however for the normal incidences ($\theta\ll 1$) which can give the dominant contribution in the current, $\chi'$ cannot be exceedingly large and is at most on the order of 1. Subsequently, we would not expect oscillatory behavior originated from $\chi'$ and as it will be clear soon only conventional AB phase $\chi$ can give rise to the magnetic oscillations known as the Fraunhofer pattern of the critical current. 
\subsection{Bound state energies and the supercurrent}\label{ABS-sub}
Now using the semiclassical framework developed above and the BdG equations, we can find the bound states energies ($|\varepsilon|<\Delta_0$) 
for any trajectory sandwiched between the interfaces. The formation of bound state can be usually understood as the result of successive Andreev reflections at the two interfaces \cite{beenakker-91}. Such Andreev bound states (ABS) are current carrying and therefore responsible for Josephson current in the short junction limit where the length is smaller than superconducting coherence length ($L\ll\xi$).
\par
In order to find the ABS energies, we invoke Eq. (\ref{evolution}) to see how electron and hole excitations evolve between the two interfaces at the presence of gauge fields. From aforementioned properties of the magnetic and strain-induced gauge fields, we can write the total gauge potentials for electron and hole excitations from different valleys in the following form,
 \begin{equation}
\begin{array}{l}
\mathbfcal{A}_{e,K}={\bf A}+{\bf A}_S,~~~~~~~~~~
\mathbfcal{A}_{e,K'}={\bf A}-{\bf A}_S,~~\\
\mathbfcal{A}_{h,K'}=-{\bf A}+{\bf A}_S,~~~~~~~
\mathbfcal{A}_{h,K}=-{\bf A}-{\bf A}_S.~~
\end{array}
\end{equation}
Subsequently, one can consider similar structure for the phase factors $\chi$ and $\chi'$ decomposed to $\chi_{B,S}$ and $\chi'_{B,S}$ where subscripts $B$ and $S$ denote magnetic field and strain-induced gauge effects, respectively. Considering electrons and holes from valleys $K$ and $K'$, respectively, the corresponding spinors at the two interfaces are connected as below,
\begin{eqnarray}\label{bq1}
 {\hat{\psi}_{e,K}|}_{R}&=e^{ 
-i(\chi_B+\chi_S)-i(\chi'_B+\chi'_S)\hat{\sigma}_z }
  \,{\hat{\psi}_{e,K}|}_{L},\\
  \label{bq2}
   {\hat{\psi}_{h,K'}|}_{R}&=e^{ 
-i(-\chi_B+\chi_S)-i(-\chi'_B+\chi'_S)\hat{\sigma}_z }
  \,{\hat{\psi}_{h,K'}|}_{L}.
  \end{eqnarray}
On the other hand, the superconducting correlations at the interfaces lead to electron-hole conversions governed by the following boundary condition \cite{titov-2006,beenakker-2008},
\begin{equation}
\hat{\psi}_{h,K'}= e^{\pm i(\frac{\phi}{2}+\beta \hat{\sigma}_x)} \hat{\psi}_{e,K}\,,~~~\beta= \arccos(\frac{\varepsilon}{\Delta_0}),
\label{bq3}
\end{equation}
at left and right interfaces, respectively. Putting Eqs. (\ref{bq1})-(\ref{bq3}) together, one can easily see that the formation of a bound state inside normal graphene region is guaranteed when,
\begin{eqnarray}
{\rm Det} &\left[ e^{i(\frac{\phi}{2}+\beta \hat{\sigma}_x)}
e^{ 
-i(\chi_B+\chi_S)-i(\chi'_B+\chi'_S)\hat{\sigma}_z }\right.\nonumber\\
&\left. -e^{
-i(-\chi_B+\chi_S)-i(-\chi'_B+\chi'_S)\hat{\sigma}_z }
e^{-i(\frac{\phi}{2}+\beta \hat{\sigma}_x)} \right]=0.
\end{eqnarray}
This results in the following relation for the ABS energies,
\begin{equation}
\label{abs}
\frac{\varepsilon}{\Delta_0}=
\sqrt{\frac{\cos^{2}(\frac{\phi}{2}-\chi_B)-\sin^{2} \chi'_S}{\cos^{2}\chi'_B-\sin^{2}\chi'_S}},
\end{equation}
as functions of phase difference $\phi$, AB phase $\chi_B$ due to the real magnetic field, and the two anomalous phases $\chi'_{B}$ and $\chi'_{S}$. Although the bound state energies have been obtained by focusing on an electron (a hole) from $K$-point ($K'$-point), one can easily check that considering the electron and hole excitations from the other valleys we will find exactly the same result. This is in agreement with the particle-hole symmetry present in superconducting heterostructures.
\par
The most interesting property of the above relation is the fact that the strain-induced AB phase $\chi_S$ does not play any role in the ABS energies and the supercurrent, subsequently. One can interpret this result due to the cancellation of strain-induced AB phases accumulated in the electron and its time-reversal hole upon propagation along the trajectory. The magnetic AB phase $\chi_B$ acts as a shift in the phase difference $\phi$ between two S parts giving rise to an effective phase difference $\tilde{\phi}_B=\phi-2\chi_B$
from which the magnetic oscillations and the famous Fraunhofer pattern originate \cite{anderson-rowell,blatter-99,ostroukh-2016}. Nevertheless, the two anomalous phases mostly impose the limitation, 
\begin{equation}
\sin^2\chi'_S\leq \cos^2(\tilde{\phi}_B/2)\leq\cos^2 \chi'_B\,,\label{limit}
\end{equation}
on the range of effective phase difference $\tilde{\phi}$ in which bound states can be formed.
A simple physical interpretation for above relation can be provided if we notice that both $\chi'_{B,S}$ cause the Dirac spinors to rotate in the $xy$ plane as given by Eqs. (\ref{bq1}) and (\ref{bq2}). Therefore, in order to have bound states, the rotation of electron and hole pseudospins during their propagation between two superconductors must be compensated. Interestingly it happens that the compensation is not possible for all effective phase differences ($\tilde{\phi}$), which subsequently gives rise to the limitation for having an ABS.
\par 
It is clear that for any trajectory ${\cal C}$ labeled by $y_0$ and $\theta$, we will get a different bound state which can be denoted by $\varepsilon_{y_0,\theta}(\phi)$ and given by Eq. (\ref{abs}). In fact, under the semiclassical framework, there exist a continuous spectrum of ABS energies depending on the vertical intercept and the slope of the trajectories. Henceforth, the contribution of each bound state in the supercurrent can be obtained from the following relation for the Josephson current density \cite{beenakker-91,moghaddam-2008}
\begin{equation}\label{deltaI}
\delta I(y_{0},\theta) = -
\frac{4e}{\hbar}\frac{d\varepsilon_{y_0,\theta}(\phi)}{d\phi} \tanh
[\frac{\varepsilon_{y_0,\theta}(\phi)}{2k_{B}T}],
\end{equation}
at a temperature $T$ ($T<T_c$). The total Josephson current can be obtained by summing over all trajectories as below,
\begin{equation}\label{totalI}
I = \frac{k_F}{2\pi}\int_{\frac{-\pi}{2}}^{\frac{\pi}{2}} d\theta  \cos\theta  \int_{0}^{W} dy_{0} \,\delta I(y_{0},\theta).
\end{equation}
In the following section, using Eqs. (\ref{phases}) and (\ref{abs})-(\ref{totalI}), we will study the effects of gauge fluxes originated from real magnetic fields or/and strains on the Josephson current in the SGS junctions.
\section{Results and discussion}\label{results}
\subsection{Supercurrent enhancement by gauge fields}\label{enhanc}
\begin{figure}[tp!] 
\centering
\includegraphics[width=0.95\linewidth]{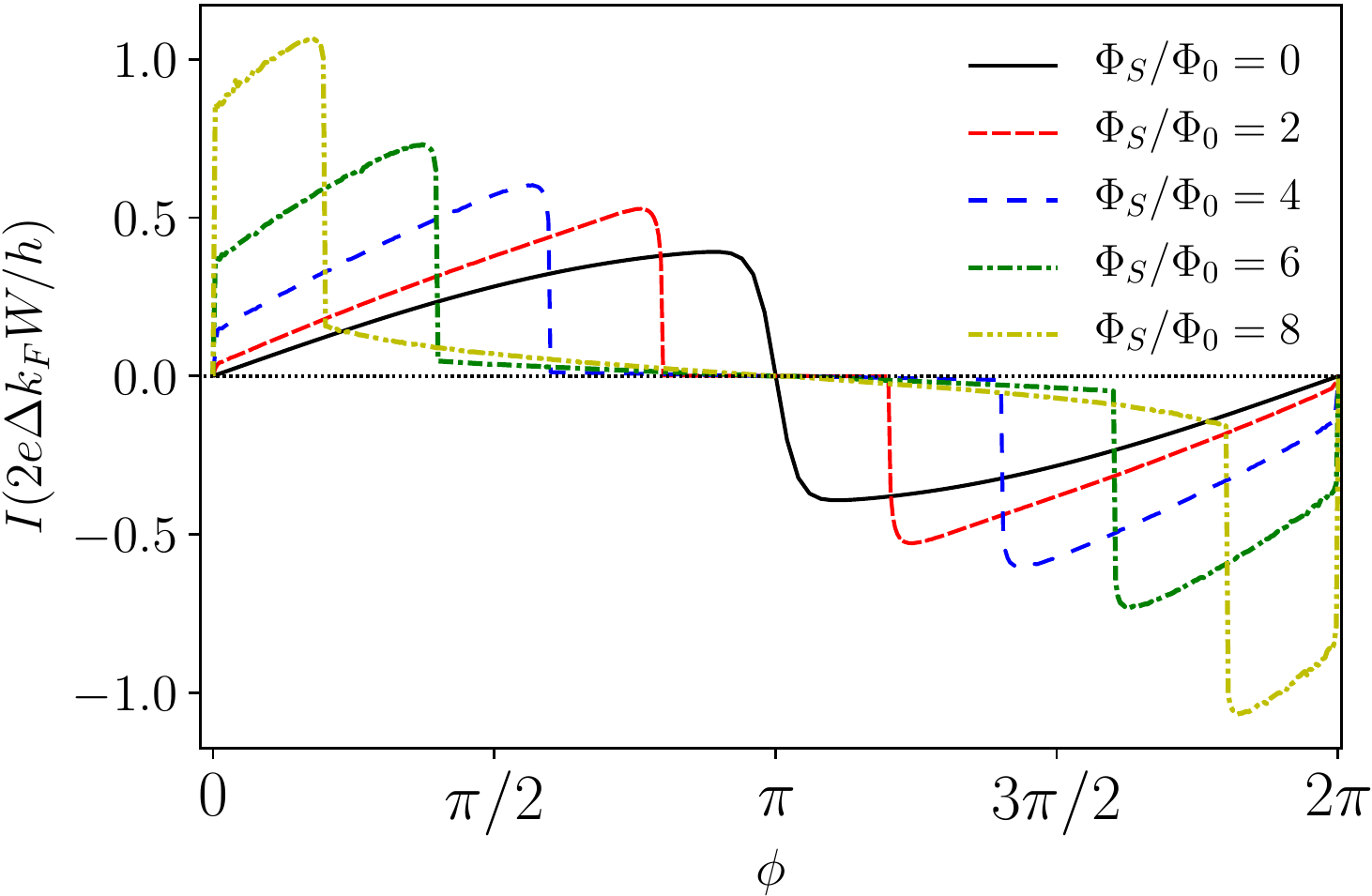}
\caption{(Color online) Current-phase relation for the SGS junction under a uniform strain-induced gauge field of various strengths given by $\Phi_S/\Phi_0$. Increasing the applied gauge field leads to larger range of phase differences in which the Josephson current is suppressed. Simultaneously, the maximum values of the Josephson current increases monotonically with the gauge field. 
The geometric aspect ratio of junction is $W/L=1$ with $k_FW=10$ and
the temperature is considered to be $T=0.05T_c$.}
\label{fig2}
\end{figure}

\begin{figure}[tp!]
\centering
\includegraphics[width=0.95\linewidth]{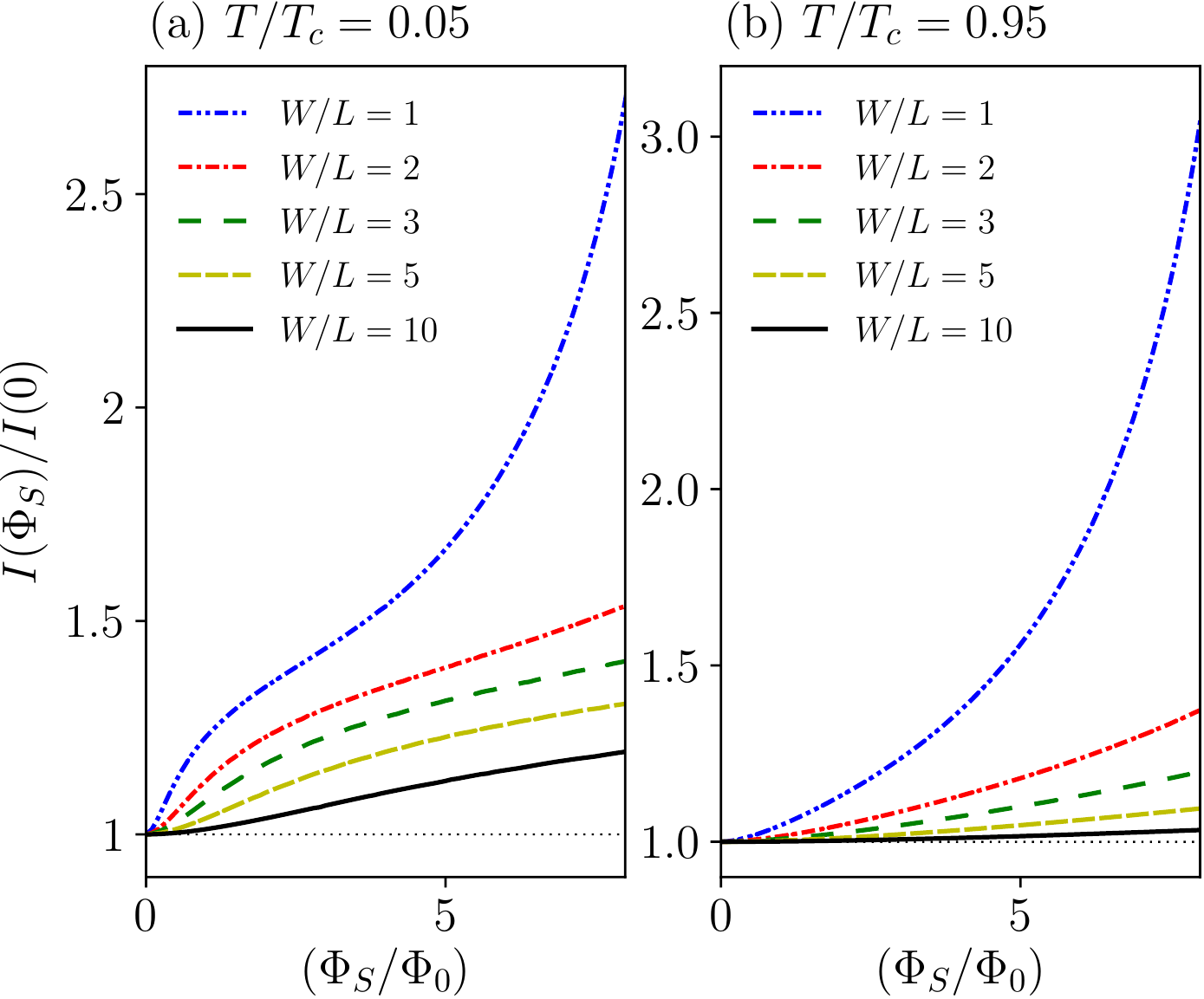}
\caption{(Color online)
Variation of critical current with the strength of strain-induced gauge field for some aspect ratios of the junction. Panels (a) and (b) correspond to the temperatures $T=0.05T_c$ and $T=0.05T_c$.
In general, the critical current monotonically increases with the gauge field, however the amplitude of variations is larger for smaller geometric aspect ratios. Here the Fermi wave vector is $k_FW=10$
}
\label{fig3}
\end{figure}

We first examine the effect of mere gauge fields 
induced by arc-shape strain on the Josephson current. It is clear that here only the phase $\chi'_S$ or the corresponding flux $\Phi_S=LWB_s$ can affect the bound state energies and the supercurrent, subsequently. 
The numerically obtained current-phase relation (CPR) for an SGS junction at the presence of a uniform gauge field is shown in Fig. \ref{fig2} for different values of $\Phi_S/\Phi_0$. 
It shows that there is a range for superconducting phase differences $\phi$ in which the Josephson current is almost suppressed. We previously in Subsec. \ref{ABS-sub} noticed that the range of phase differences over which a abound state can exist is constrained due to the rotation of pseudospin caused by anomalous phases $\chi'$. This results in a constraint on the range of $\phi$'s with finite Josephson current which can be approximately obtained from Eqs. (\ref{phases}) and (\ref{limit}) as,
\begin{equation}
\pi-\frac{\pi}{2k_FW}\frac{\Phi_S}{\Phi_0}<\phi<\pi+\frac{\pi}{2k_FW}\frac{\Phi_S}{\Phi_0},
\end{equation}
by considering only the normally incident trajectories ($\theta=0$) which have the dominant contribution in the supercurrent. 
As a consequence of confined range of $\phi$'s to have ABS, by increasing $\Phi_S/\Phi_0$ the bound states energies vary sharply with the phase difference and therefore the Josephson current can reach higher values as it can be seen in Fig. \ref{fig2}. Then it follows that the critical current must be raised by increasing the strength of strain and corresponding gauge field.
This is in contrast to the Fraunhofer-type behavior induced by real magnetic fields as it is shown in Fig. \ref{fig3} for various aspect ratios of the junction and at two different temperatures. Interestingly, the effect of gauge field is much stronger for longer junction with smaller values of $W/L$ as can be understood from the dependence of gauge-induced anomalous phase $\chi'$ on the length followed from Eq. (\ref{chi-chip}).
\subsection{Josephson vortices and Fraunhofer patterns}
\begin{figure*}[htp!]
\centering
\includegraphics[width=0.95\linewidth]{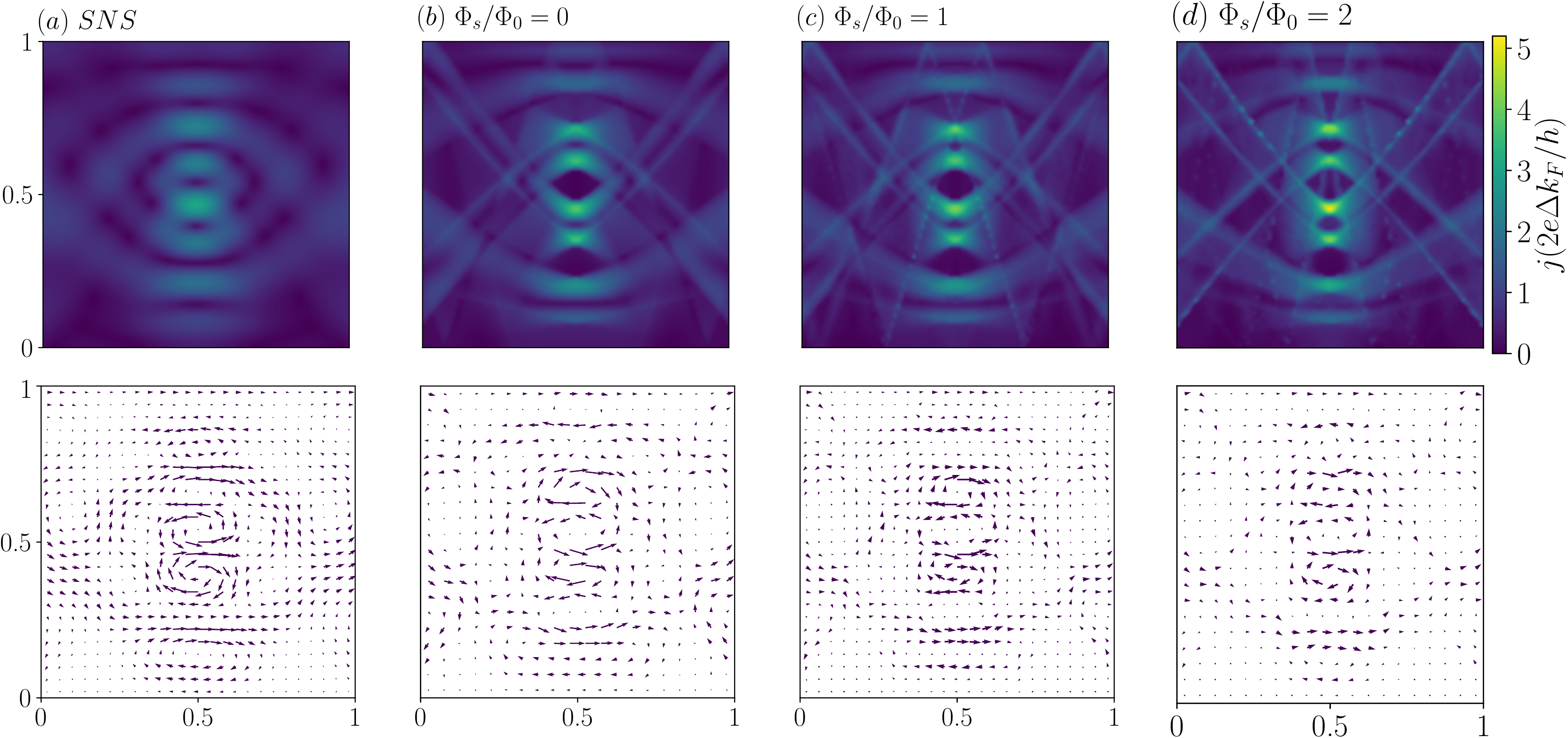}
\caption{(Color online) Formation of Josephson vortices at the presence of magnetic flux $\Phi_B/\Phi_0=4$ in (a) an SNS junction and (b)-(d) SGS junctions subjected to different values of strain-induced gauge fields. Each column corresponds to the same situation with upper and lower panel indicating the intensity plot and the vector field of the local current density, respectively. While the generic form of the vortex pattern is remained almost the same in all cases, but the local current profile in SGS junctions has a more localized pattern. The localization is intensified at the presence and by increase of the gauge field. The parameters considered in all subplots are $k_FW=10$, $W/L=1$, $\phi=\pi/4$, and $T/T_c=0.95$.}
\label{fig4}
\end{figure*}

In this part, we would like to investigate the supercurrent density profiles in the presence of magnetic and pseudomagnetic fields. 
As it has been mentioned in the introduction, when the normal region of an SNS junction is subjected to a uniform magnetic field, the supercurrent profile is strongly modulated due to the quantum interferences which lead to the vortex-like circular flow patterns. In the semiclassical picture, the local current density at each point can be obtained by summing over the contributions of all trajectories passing through that point or equivalently by mere integration over $\theta$. 
Figure \ref{fig4} shows the supercurrent profile of a conventional SNS junction and SGS junctions in the presence of a gauge field 
of various strengths $\Phi_S/\Phi_0=0,1,2$, all subjected to a constant magnetic field flux $\Phi_B/\Phi_0=4$ and a phase difference $\phi=\pi/4$ between two S electrodes. 
In all cases series of vortices and antivortices are seen which are mainly located at the middle of the junction. Furthermore, we see that compared to an SNS junction, SGS case has more profound and localized vortex structure. Intriguingly, when the strain is imposed to the graphene, the local Josephson current is further localized by increasing the strength of the gauge field. 
\par
It is very instructive to discuss how vortex patterns are formed in Josephson current density in a generic SNS system,
before providing detailed discussion about the SGS junction and the role of strain-induced gauge fields. To this end, let us first examine the current density over the vertical line exactly at the middle of the junction ($x=L/2$). One can easily see that the current flows in the horizontal direction, there. The reason comes from the mirror symmetry between the trajectories with opposite angles $\pm\theta$ passing from the line $x=L/2$, which subsequently results in the same AB phases and differential current amplitudes $\delta I(\theta)=\delta I(-\theta)$ (One should remember that the AB phase $\chi$ is proportional to the area underneath of the trajectory). In addition, all straight trajectories crossing the point $(x=L/2,y)$ enclose the same area below themselves and as a result, their contributions to the local current add up coherently. Then moving along the vertical direction, the AB phase $\chi_B$ linearly increases with $y$ for most of the trajectories. Therefore one can conclude that the bound state energies and the Josephson current density vary almost periodically along $y$-direction according to Eqs. (\ref{abs}) and (\ref{deltaI}). Deviation from perfect periodic variations becomes visible close to the horizontal boundaries of the junction where a substantial part of the trajectories is reflected once or more at the edges. In contrast to the straight paths, the zigzag ones passing from a certain point in the middle of the junction can have different areas below them which give rise to different AB phases as well. So we would expect a weak aperiodicity which can be moderate close to the horizontal edges as seen in Fig. \ref{fig4}(a).  Away from the middle of the junction, $\delta I(\theta)$ will be no longer symmetric with respect to the angle because of the difference in the areas $S(y_0,\pm\theta)$. Therefore, the local current at points $x\neq L/2$ after integration over $\theta$ can have a vertical component which qualitatively explains the appearance of the circular flow pattern around certain points mostly located at $x=L/2$. Moreover, besides the major line of the vortex-antivortex series, very weak vortex flows can be found far from the middle of the junction and close to the SG interfaces.
\par
When we turn to the graphene-based Josephson junctions, at the presence of a real magnetic field, anomalous phase $\chi'_B$ will come into play besides the AB phase.
We have already seen in previous parts that the anomalous phases leads to the pseudospin rotation which subsequently puts some constraints on the range of parameters over which ABS can be formed. So for any constant phase difference $\phi$, due to the constraint given by Eq. (\ref{limit}) some trajectories could not give rise to ABS in contrast to the SNS junction. Moreover, on the same ground as we discussed the underlying physics of Josephson current enhancement by the gauge fields, the trajectories hosting an ABS give rise to larger Josephson current. Putting all together, we can understand why the profile of the local supercurrent density is more localized and have larger peaks in graphene-based Josephson junctions as seen in Fig. \ref{fig4}(b) compared to an SNS junction. On the same ground, when strain is applied to graphene, the corresponding anomalous phase $\chi'_S$ leads to further localization of the local current flow patterns. In fact as it is seen from Figs. \ref{fig4}(c),(d) the interference pattern is not drastically influenced by the pseudomagnetic fields as long as the their strength is not strong compared to the real magnetic field. So when $\Phi_S\lesssim \Phi_B$, the major effect of pseudomagnetic fields is that the spatial variations becomes increasingly sharper at the presence of a finite $\Phi_S$. This is again along with the fact that strain-induced gauge fields have not any AB type effect which can result in further modulation of the supercurrent. 
\begin{figure}[t!]
\centering
\includegraphics[width=0.95\linewidth]{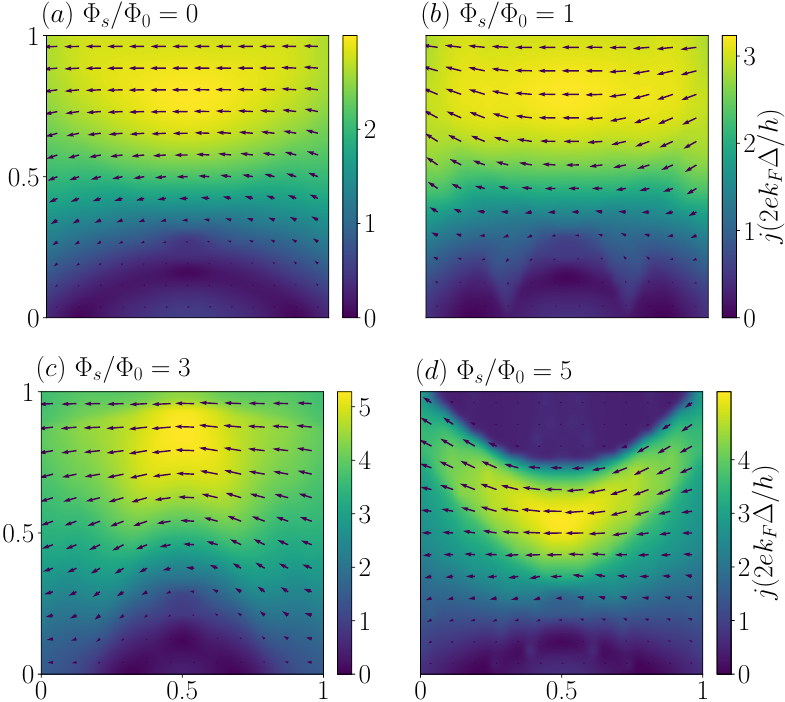}
\caption{(Color online)
Evolution of local current density profile with pseudomagnetic field at the presence of a constant magnetic flux $\Phi_B/\Phi_0=0.5$.
While the trend from (a) to (c) indicates that only local current becomes more localized and the maximum values increase, the behavior in the presence of large gauge field ($\Phi_S/\Phi_0=5$) is drastically different. In the last situation a big halo appears on the top of the junction where the supercurrent is almost diminished. The halo region has sharp boundaries with its surroundings in contrast to the conventional vortex cores (like the one on the lower edge) where the current density varies smoothly with the distance from the vortex center. The other parameter used here are the same as those in Fig. \ref{fig4}.}
\label{fig5}
\end{figure}
\par
When the gauge fields originated from the strain are strong enough in comparison to the the applied magnetic field, another interesting phenomenon shows up.  This new effect is the appearance of a finite region where the current is almost suppressed due to the gauge fields, as clearly can be seen in Fig. \ref{fig5}(d) where a large pseudomagnetic flux $\Phi_S/\Phi_0=5$ is imposed to the junction. Such behavior is essentially different from the main trend in Figs. \ref{fig5}(a)-(c) corresponding to $\Phi_S/\Phi_B=0,1,3$ in which the supercurrent only becomes slightly squeezed by pseudomagnetic field in accordance with the above discussion. A big difference of the new region with conventional vortex cores is that it has much sharper boundary with the region of finite supercurrent as can be seen from Fig. \ref{fig5}(d). Interestingly the halo region of supercurrent suppression induced by large pseudomagnetic field is placed on top of the junction which have the largest supercurrent when $\Phi_S=0$. The origin of new vortex cores is indeed nothing but the constraint (\ref{limit}) in which by increasing anomalous phase $\chi'_S$ from $0$ to $\pi/2$ the range of effective phase differences $\tilde{\phi}_B=\phi-2\chi_B$ with contribution in the current decreases. Therefore, for a constant phase difference $\phi$, the AB phase $\chi_B$ or equivalently the trajectories with nonvanishing current will be constrained. To better illustrate this effect let us consider the normally incident trajectories ($\theta=0$) for which the AB phase is simply $\chi_B=\pi(\Phi_B/\Phi_0)(y/W)$. Then we can rewrite the constraint due to the strain-induced anomalous phase as below,
\begin{equation}
|\cos(\frac{\phi}{2}-\pi\frac{\Phi_B}{\Phi_0}\frac{y}{W})|\geq \sin(\frac{\pi}{2 k_F W}\frac{\Phi_S}{\Phi_0})
\end{equation}
Invoking the parameters used in Fig. \ref{fig5}(d), one can easily see that for $y/W>3/4$ the above condition is not satisfied. Therefore all trajectories with $\theta=0$ and $y_0>3/4$ do not carry a supercurrent which suggest by itself that there must be a region of strong supercurrent suppression around the upper edge of the junction. By more detailed analysis of the other trajectories one can obtain the precise boundary of the halo region with the surrounding in Fig. \ref{fig5}(d).
\begin{figure}[t!]
\centering
\includegraphics[width=0.8\linewidth]{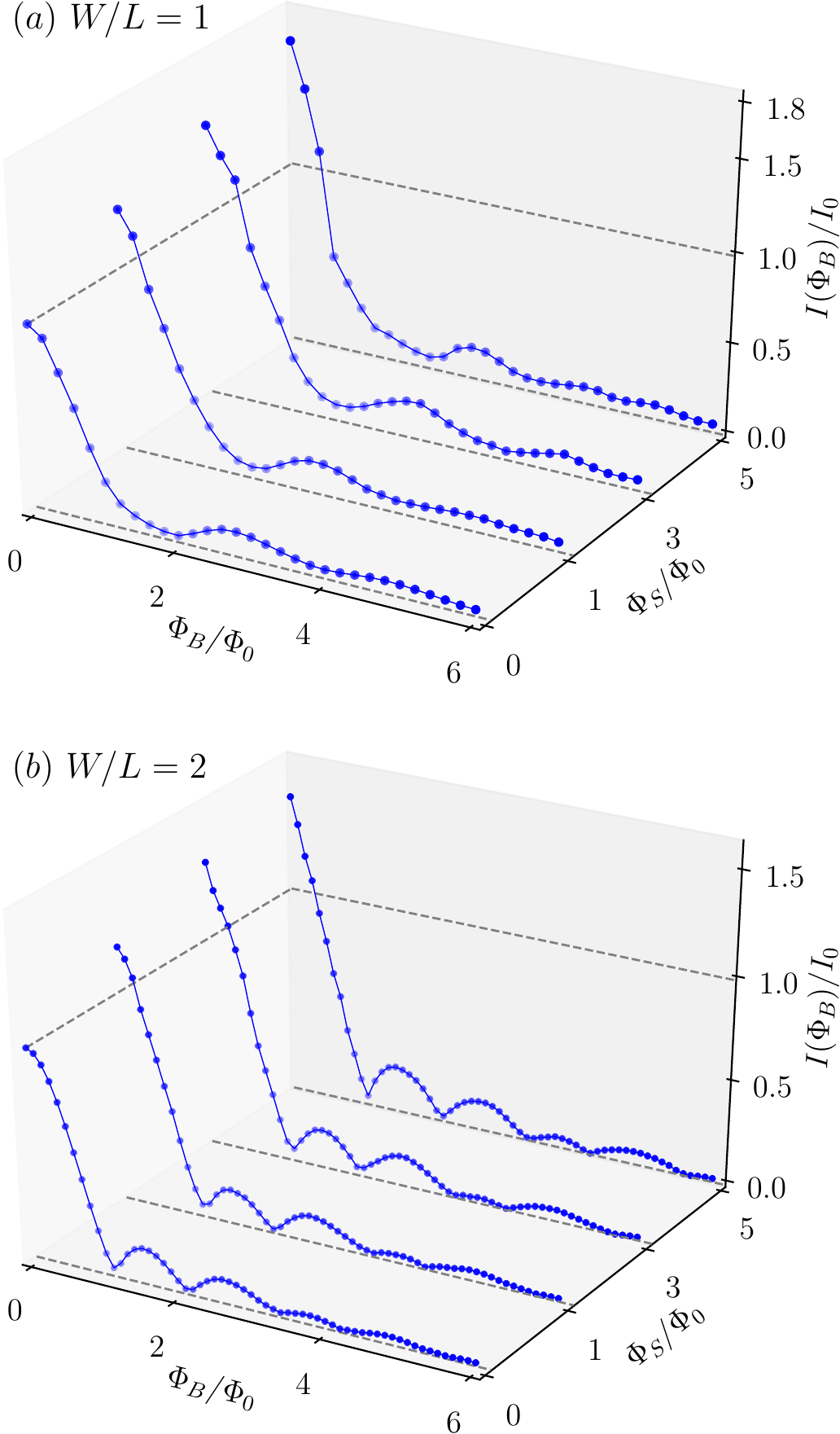}
\caption{(Color online) Magnetic oscillations of the critical Josephson current known as Fraunhofer pattern when the graphene junction is subjected to the pseudomagnetic fields of different strengths $\Phi_S/\Phi_0=0,1,3,5$. (a) and (b) correspond to different width to length ratios of the junction $W/L=1,2$, respectively. While the maximum value of the critical current at $\Phi_B=0$ increases monotonically with the pseudomagnetic field, its effect is less apparent at larger magnetic fluxes. By increasing the width of the junction, oscillations become more regular while aperiodic behavior appears when $W\lesssim L$.
}
\label{fig7}
\end{figure}
\par
At the end of this part, we examine evolution of the Fraunhofer patterns with the pseudomagnetic field as depicted in
Fig. \ref{fig7}. In the absence of gauge field ($\Phi_S=0$), the conventional magnetic oscillations of the critical current are retained. Particularly at the limit of wide junctions $W/L\gg 1$, the period of oscillations $\delta\Phi_B$ is $\Phi_0$ while for narrower junctions with $W/L\lesssim1$, it switches to $2\Phi_0$ in agreement with previous investigations in SGS and SNS systems \cite{cserti-2016,blatter-99}. When a constant pseudomagnetic or gauge field is present, the overall behavior of the Fraunhofer pattern remains almost the same as when it is absent except for increase in the maximum current at $\Phi_B=0$. This result again originates from the fact that gauge fields do not cause any AB type oscillation or modulation of the supercurrent. Returning back to the discussion over the supercurrent density profiles, when the magnetic field was strong enough to modulate the current and induce couples of vortices inside the junction, the pseudomagnetic field leads to squeezed profiles with sharper peaks. So when we look at the whole critical current passing through the junction, these details are washed out by integration over the local current density, a fact that justifies why the Fraunhofer pattern is not strongly influenced by gauge fields.
In other words, as it has been already seen in our other results in part \ref{enhanc}, Fig. \ref{fig3} and Fig. \ref{fig5}, the effect of gauge fields is very strong either in the absence of real magnetic fields or when the real magnetic flux is small namely $\Phi_B/\Phi_0\lesssim 1$.
 
\subsection{Discussion and experimental relevance}
As it has been mentioned in the introduction the emergence of gauge fields due to the strain is a result of its 2D nature as well as Dirac character of the quasiparticles.
While many theoretical and experimental investigations have been carried out to understand and exploit the strain-induced gauge fields for possible applications, only few works focused on the interference phenomena especially AB effect due to the gauge fields. Among them an interesting proposal made by de Jaun et al. \cite{juan-natphys} to observe interference effects due to a local deformation in the electronic propagation encircling the strained region. Here we have focused on the graphene-based superconducting heterostructures and investigated the gauge fields effects on the Josephson current, its density profile and magnetic oscillation pattern. As it has been shown above, the critical Josephson current as well as the local supercurrent density are qualitatively influenced by presence of strain-induced pseudomagnetic fields. It should be mentioned that the fact that ABS are formed from time-reversal partners introduces a fundamental difference between superconducting system we are studying and AB interferometry scenario in Ref. \cite{juan-natphys}. Here the AB phases attributed to the gauge field does not enter in the supercurrent at all and only the anomalous phase factor $\chi'_S$ which corresponds to pseudospin rotation governs the underlying physics of our findings. Of particular importance, in SGS junction, the Josephson current is unexpectedly enhanced by strain-induced gauge fields. 
\par
All the predicted features can be possibly tested in future experiments provided by the validity of some assumptions in the current work. First of all, the excitations should keep their phase coherence while propagating throughout the junction. However this is not an elusive situation to current experimental devices and phase coherence lengths in SGS systems could be large enough compared to the junction size \cite{vandersypen-2015,benshalom-2015,yacoby-natphys,yacoby-2017}. On the other hand, one may truly doubt the validity of our results at the presence of strong inter-valley scatterings which can mix the electrons (or holes) from the two valleys. But nowadays very clean samples of suspended graphene can be manufactures with weak or moderate inter-valley scatterings \cite{gorbachev2014,schaibley2016}. Therefore the overall experimental feasibility of our theoretical study is quite high even for the currently available setups.   

\section{Conclusions}\label{conc}
In this work, we have found anomalous quantum interference effects in graphene-based Josephson junctions subjected to the strain-induced pseudomagnetic fields  besides real magnetic fields. As a key finding, it has been shown that the Josephson current is enhanced by applying an arc-shape strain which gives rise to a constant pseudomagnetic or gauge field inside the graphene region sandwiched between superconductors. Such behavior is in complete contrast to the well-known Fraunhofer pattern of the critical current with an oscillatory decaying dependence on the the magnetic flux. The fundamental difference between real magnetic fields and strain-induced gauge fields regarding their symmetries under time-reversal has been proven to be responsible for very different effects of magnetic and pseudomagnetic fluxes. We further have studied the combined effects of both type of fields simultaneously applied to the graphene Josephson junction. While the magnetic fields leads to the formation of Josephson vortices and Fraunhofer pattern, the presence of gauge fields result in squeezing of the local current density in circular flow patterns. Moreover, it has been revealed that stronger gauge fields compared to the magnetic fields, can affect the supercurrent flow pattern more crucially by forming large vortex cores with almost vanishing current in a finite region. So, our investigation have revealed unexpected features of the interference phenomena in graphene-based Josephson junction due to the strain-induced gauge fields which trigger further theoretical and experimental studies on the physics of gauge fields in graphene/superconductor hybrid structures.

\section*{Acknowledgements}
A. G. M. acknowledges financial support from the Iran
Science Elites Federation under Grant No. 11/66332.

\appendix
\section{Calculation of area enclosed by trajectories}
\label{appendix}
Here we calculate the Aharonov-Bohm phase of each trajectory and give the exact form of the corresponding area $S_{y_0,\theta}$. By choosing the Landau gauge as 
$\mathbfcal{A}={\cal B} y {\bf x}$ it is clear that the line integral of gauge potential along the vertical NS interfaces as well as the lower edge of the graphene vanishes. So the contribution of the 
trajectory ${\cal C}$ in the line integral is the same as the corresponding closed loop and can be related to the area underneath ${\cal C}$ via Stokes' theorem,
\begin{equation}
\int_{\cal C}    d\ell {\bf n}\cdot {\mathbfcal A}={\cal B} S_{y_0,\theta}.
\end{equation}
Now in order to calculate $S_{y_0,\theta}$, for a given angle $\theta$ and offset $y_0$, we first determine the position and the number of reflection points of the trajectory at the two edges. For the sake of simplicity we divide the discussion to two parts for positive and negative angles. 
When $0 < \theta < \pi/2$ we have, 
\begin{eqnarray}
&& N= [\frac{y_{0} +L \tan\theta}{W}],\\
&& x_n= (nW-y_{0})\cot\theta,~~~~ 1 \leq n \leq N,
\end{eqnarray}
respectively. Then by simple geometry we can obtain the area for three different cases of $N=0$, an even $N$, and an odd $N$, as the following, respectively, 
\begin{eqnarray}
&& S_{y_0,\theta} = L  y_{0} +\frac{L^{2}}{2}\tan\theta,\\
&& S_{y_0,\theta} = y_{0} \frac{x_{1}}{2} +W \frac{x_{N}}{2}+\dfrac{(L-x_{N})^{2}}{2}\tan\theta,\\
&& S_{y_0,\theta}=y_{0} \frac{x_{1}}{2}+W (L-\frac{x_N}{2}) -\frac{(L-x_{N})^{2}}{2}\tan\theta.
\end{eqnarray}

Similarly for negative angles ($-\pi/2 <\theta\leq0$) we can obtain the number of reflections and their $x$ positions which reas,
\begin{eqnarray}
&& N= [\frac{W- y_{0} + L |\tan\theta|}{W}],\\
&& x_n= [(n-1)W+y_{0}]|\cot\theta|,~~~~ 1 \leq n \leq N,
\end{eqnarray}
Subsequently, the area for $N=0$, an even $N$, and an odd $N$ when $\theta$ is negative angles can be obtained as, 
\begin{eqnarray}
&& S_{y_0,\theta} = L  y_{0} -\frac{L^{2}}{2}|\tan\theta|,\\
&& S_{y_0,\theta} = (y_{0}-W) \frac{x_{1}}{2} +W \frac{x_{N}}{2}\nonumber\\
&&~~~~~~~~~~~~~~~~~~~~~~~~~~+\frac{(L-x_{N})^{2}}{2}|\tan\theta|,\\
&& S_{y_0,\theta} = (y_{0}-W) \frac{x_{1}}{2} +W (L-\frac{x_{N}}{2})\nonumber\\
&&~~~~~~~~~~~~~~~~~~~~~~~~~~-\frac{(L-x_{N})^{2}}{2}|\tan\theta|,
\end{eqnarray}
respectively.

\section*{References}
\bibliography{hkfrn.bib}
\end{document}